\documentclass[prb,
superscriptaddress,showpacs,amsmath,amssymb]{revtex4}

\usepackage{graphicx}

\begin{document}

\title{Condensation of fermion zero modes in the vortex in nodal superfluids/superconductors}

\author{G.E.~Volovik}
\affiliation{Low Temperature Laboratory, Department of Applied Physics, Aalto University, PO Box 15100, FI-00076 AALTO, Finland}
\affiliation{Landau Institute for Theoretical Physics, acad. Semyonov av., 1a, 142432,
Chernogolovka, Russia}
\affiliation{
Nordita,
KTH Royal Institute of Technology and Stockholm University,
Roslagstullsbacken 23, SE-106 91 Stockholm, Sweden}

\date{\today}

\begin{abstract}
The energy levels of the fermions bound to the vortex are
considered for vortices in the superfluid/superconducting systems which contain  the symmetry protected plane of zeroes in the gap function in bulk. The Caroli-de Gennes-Matricon branches with different angular momentum quantum number $n$ approach zero energy level at $p_z\rightarrow 0$. Such condensation of the energy levels is the consequence of the bulk-vortex correspondence in topological superfluids/superconductors. In a given case  this is the connection  between the Dirac line of zeroes in the bulk spectrum and the level condensation in the vortex core.
The density of states of the bound fermions diverges at zero energy giving rise to the $\sqrt{\Omega}$ dependence of DoS in the polar phase of superfluid $^3$He rotating with the angular velocity $\Omega$ and to the $\sqrt{B}$ dependence of DoS for superconductors
in the $(d_{xz} + i d_{yz})$-wave pairing state.
\end{abstract} 
\

\


\maketitle

\begin{figure}
 \includegraphics[width=0.8\textwidth]{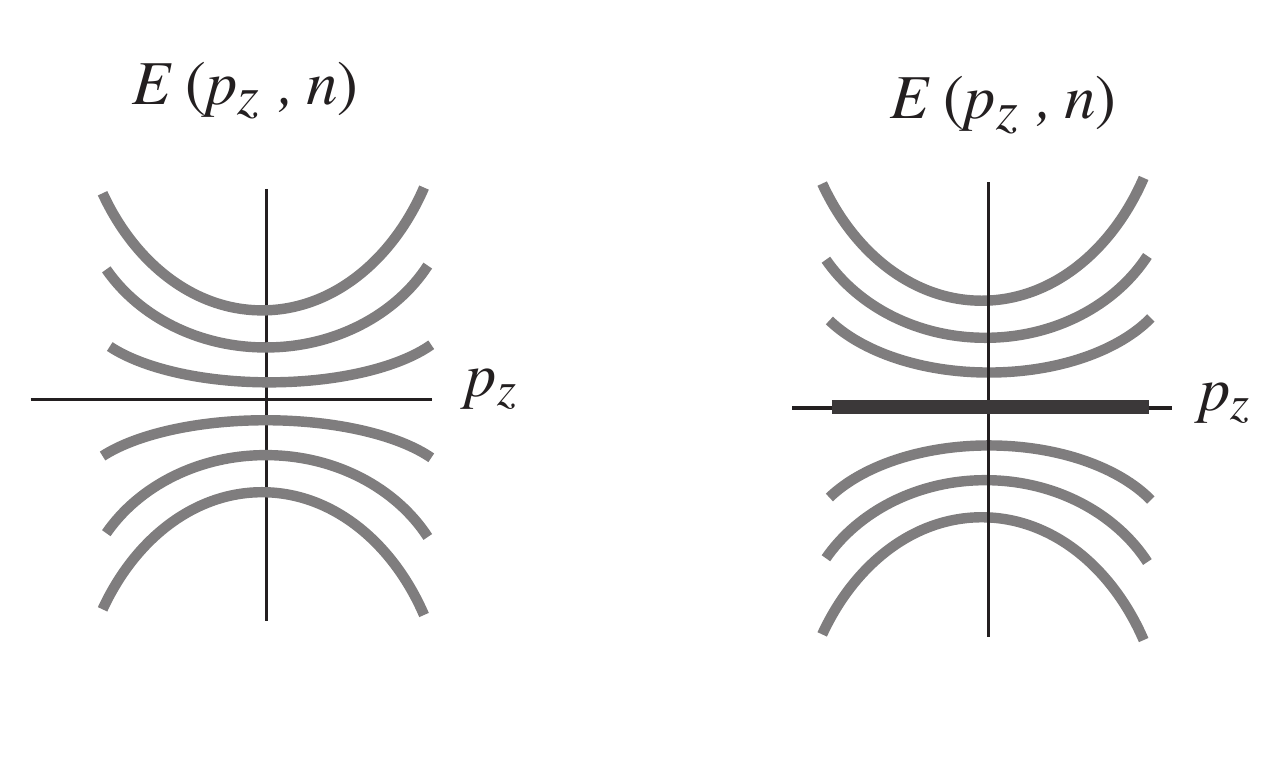}
 \caption{ Illustration of the spectrum of fermion bound states on symmetric singly-quantized ($m=\pm 1)$ vortices.
 ({\it Left}): bound states in the nontopological $s$-wave superconductors and in the system with even chiralty number $N=2k$. ({\it Right}): bound states on the most symmetric vortex in the chiral Weyl superfluid $^3$He-A. The spectrum contains the flat band, which terminates on the projections of two Weyl points in bulk to the vortex line.\cite{Volovik2011} The flat band appears also in other systems with odd chiralty number $N=2k+1$.
 }
 \label{CoreStates}
\end{figure}

The spectrum of the low-energy bound states in the
core
of the symmetric vortex with winding number $m=\pm 1$ in the isotropic
model of
$s$-wave superconductor was obtained in a microscopic (BCS) theory by Caroli,
de
Gennes and Matricon,
\cite{Caroli} see Fig. \ref{CoreStates} ({\it left}):
\begin{equation}
E_n=\left(n+{1\over 2}\right)\omega_0(p_z) \,.
 \label{Caroli}
\end{equation}
Here $p_z$ is the momentum of the bound states along the vortex line, and $n$ is related to the angular momentum quantum number $L_z$.
This spectrum is two-fold degenerate due to spin degrees of freedom.
The level spacing -- the so called minigap -- is small compared to the energy gap of the
quasiparticles
outside the core, $\omega_0\sim \Delta^2/E_F\ll\Delta$. 

For the chiral superfluid/superconductor with an odd winding number of the phase of the gap function  
in momentum space (i.e. $\Delta({\bf p}) \propto (p_x + ip_y)^N$ with odd $N$), the spectrum of fermions in the symmetric vortex is modified, \cite{Volovik1999,Volovik2003}  see Fig. \ref{CoreStates} ({\it right})
for the most symmetric vortex in the Weyl superfluid $^3$He-A:
\begin{equation}
E_n=n\omega_0(p_z)\,.
 \label{chiral}
\end{equation}
The spectrum contains the zero energy states at $n=0$. In 
the two-dimensional case the $n=0$ levels represent 
two Majorana modes.\cite{ReadGreen2000,Ivanov2001} 
The 2D half-quantum vortex, which is the vortex in one spin component, contains single Majorana mode.
In the 3D case the Eq.(\ref{chiral}) at $n=0$ 
describes the flat band in the spectrum of the bound states:\cite{KopninSalomaa1991} all the states in the interval $-p_0<p_z<p_0$ have zero energy, where $p_0\hat{\bf z}$ and $-p_0\hat{\bf z}$
mark the positions of two Weyl points in the bulk material. \cite{Volovik2011}
The connection between the positions of the Weyl points in bulk, and the boundary of the flat band in the vortex core is the consequence of the bulk-vortex correspondence in topological materials.
For numerical simulation of the flat band in the core of the $^3$He-A vortex see
Ref.\cite{PO-YAO-CHANG2015}.

The $n=0$  level with the flat band appears also in spin-singlet superconductors with $(d_{xz} + i d_{yz})$-wave pairing.\cite{LeeSchnyder2015} This is because  the gap function  $\Delta({\bf p}) \propto p_z(p_x + ip_y)$ has an odd chirality number $N=1$ and the Weyl points in bulk. Such pairing has been suggested in the heavy-fermion compound URu$_2$Si$_2$.\cite{Mizushima2016,Kittaka2016} The $n=0$  level with the flat band does not appear in the core of superconductors with $(d_{x^2-y^2} + i d_{xy})$-wave pairing,\cite{LeeSchnyder2015}   where the gap $\Delta({\bf p}) \propto (p_x + ip_y)^2$ has $N=2$, and the spectrum contains the Weyl points with double topological charge and quadratic dispersion.\cite{Volovik2003}

Here we consider vortices, in which the minigap $\omega_0(p_z)$ vanishes at $p_z=0$. This leads to the enhanced density of states of the fermions in the vortex core, and as a consequence to the non-analytic behavior 
of the DoS as a function of magnetic field in superconductor or of rotation velocity in superfluid.

Examples are provided by vortices\cite{Autti2015}  in the recently discovered\cite{Dmitriev2015b} non-chiral ($N=0$) spin-triplet polar phase of superfluid $^3$He, and by vortices in superconductors where the pairing state has the symmetry protected plane of gap zeroes,\cite{VolovikGorkov1985}
such as the chiral ($N=1$) spin-singlet superconductor with $(d_{xz} + i d_{yz})$ pairing.\cite{LeeSchnyder2015}
We start with the polar phase, whose order parameter matrix is
\begin{eqnarray}
 \hat\Delta_{\bf p}=\sigma_z\Delta({\bf r}) \frac{p_z}{p_F} \,.
 \label{OrderParameterTriplet}
\end{eqnarray}
 Here $\sigma_z$ is the
Pauli matrix for conventional spin. The polar phase contains topologically protected Dirac line in the spectrum, at $p_z=0$ and $p=p_F$, which gives rise to the flat band of the surface fermions according to the bulk-surface correspondence, see reviews \cite{VolovikSilaev2014,SchnyderBrydon2015}.
Dirac lines exist in cuprate superconductors\cite{Volovik1993} and also in semimetals.\cite{HeikkilaVolovik2011,HeikkilaVolovik2015,Mikitik2006,Mikitik2008,Weng2014,Xie2015,Kane2015,Yu2015}

We consider the most symmetric vortices, i.e. without extra symmetry breaking in the vortex core. 
For vortces with broken discrete symmetries the bulk-vortex correspondence is violated, and the character of the spectrum 
changes, see the spectrum of bound states on vortices with broken symmetries in the topological superfluid  $^3$He-B, which is time reversal invariant in bulk.\cite{Silaev2009,Silaev2015} 
The symmetric singly quantized vortex has the following structure of the
order parameter in the core:
\begin{equation}
\Delta({\bf r}) =\Delta(r) e^{im\phi}~, ~ \Delta(r\rightarrow \infty) =\Delta_0\,,
\label{OrderParameterInVortex}
\end{equation}
where $z$, $r$, $\phi$ are the coordinates of the cylindrical system
with
the axis $z$ along the vortex line; $m=\pm 1$; $\Delta_0$ is the gap amplitude far from the vortex core.
In the half quantum vortex observed in Ref.\cite{Autti2015}  only single spin component has vorticity:
$m=\pm 1$ for, say, $S_z=+1$ and $m=0$ for $S_z=-1$.

Using standard procedure \cite{Volovik1999}
one obtains the minigap in the polar phase:
\begin{equation}
\omega_0(p_z)  =\frac {
\int_{0}^\infty dr |\psi_0(r)|^2 
\frac{\Delta(r)|p_z|}{p_Fqr }
}
{\int_0^\infty dr
|\psi_0(r)|^2}
~~\,,~~q=\sqrt{p_F^2-p_z^2}\,,
 \label{QuasiclasicalEnergy}
\end{equation}
where
\begin{equation}
\psi_0(r)=
\exp{
\left(-\int^r_0 dr' 
\frac{\Delta(r')|p_z|}{p_F v_F}\right)}
\,,
 \label{WaveFunction}
\end{equation}
is the wave function of the bound state.
\begin{figure}
 \includegraphics[width=0.8\textwidth]{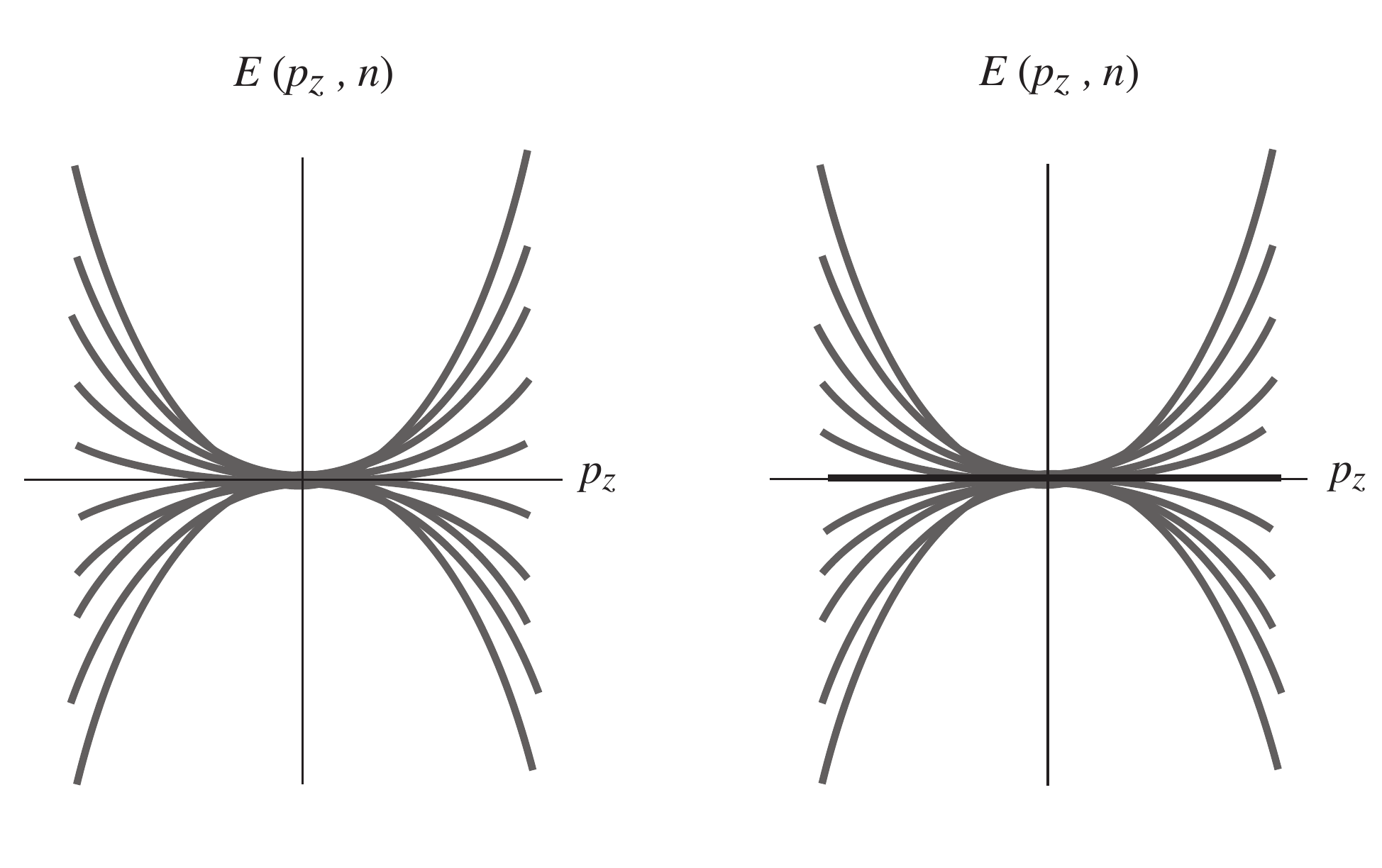}
 \caption{({\it top}): Illustration of the spectrum of fermion zero modes at $|p_z|\ll p_F$ on vortices in the polar phase of superfluid $^3$He ({\it left}) and in  the chiral $(d_{xz} + i d_{yz})$-wave superconductor  ({\it right}). The branches with different $n$ approach zero-energy level at $p_z\rightarrow 0$.
In addition, the vortex in $(d_{xz} + i d_{yz})$-wave superconductor contains the flat band at $n=0$.\cite{LeeSchnyder2015}
 }
 \label{VortexFermions}
\end{figure}

For small $p_z\ll p_F$ the minigap in the polar phase has the following form:
\begin{equation}
 \omega_0(p_z) =  \omega_{00}
\frac{p_z^2}{p_F^2}\ln\frac{p_F^2}{p_z^2} ~~,~~ \omega_{00}\sim \frac{\Delta_0^2}{E_F}
\,,
 \label{spectrum}
\end{equation}
where $\omega_{00}$ has an order of the minigap in the conventional $s$-wave superconductors.
The spectrum is shown in Fig. \ref{VortexFermions}. All the branches with different $n$ touch the zero energy level. It looks as the flat band in terms of $n$ for $p_z=0$. This is the consequence of the Dirac nodal line in bulk at $p_z=0$, which is another manifestation of the bulk-vortex correspondence. 

At $p_z\rightarrow 0$ the size of the bound state  wave function diverges and becomes larger than the core size, but still the spectrum corresponds to the bound state, since it is within the bulk gap at fixed $p_z$ if $n\omega_{00}
p_z^2/p_F^2< \Delta_0 |p_z|/p_F$.  The critical size of the bound state  wave function, at which the consideration fails, is the inter-vortex distance in the vortex lattice, when the wave functions on neighboring vortices overlap.

The effect of squeezing of all energy levels $n$ towards the zero energy
at $p_z\rightarrow 0$ can be called the condensation of Andreev-Majorana
fermions in the vortex core. 
The condensation leads to the divergent density of states (DoS) at small energy.  In the vortex cluster with the vortex density density $n_V$ the DoS is
\begin{equation}
N_V=n_V\sum_n \int \frac{dp_z}{2\pi} \delta(\omega - (n+1/2) \omega_0(p_z) ) \,.
 \label{DOS}
\end{equation}
In calculation of Eq.(\ref{DOS}) we assume that the relevant values of $n$ are large, and instead of summation over $n$ one can use the integration over $dn$:
\begin{equation}
N_V = n_V \int \frac{dp_z}{2\pi} \frac{1}{\omega_0(p_z)}\,.
 \label{DOS2}
\end{equation}
According to Eq.(\ref{spectrum}) the integral in Eq.(\ref{DOS2}) diverges at small $p_z$. The infrared cut-off is provided by the intervortex distance $r_V=n_V^{-1/2}$: the size of the wave function of the bound state $\xi p_F/|p_z|$ approaches the intervortex distance when  $|p_z|\sim p_F\xi/r_V$. This cut-off leads to the following dependence of DoS on the intervortex distance:
\begin{equation}
N_V \sim \frac{p_F^2}{\Delta_0r_V}\,.
 \label{DOS3}
\end{equation}
The result in Eq.(\ref{DOS3}) is by the factor $r_V/\xi$ larger than the DoS of fermions bound to conventional vortices.  Since in the vortex array 
$r_V \propto \Omega^{-1/2}$, the DoS has the non-analytic dependence on rotation velocity, $N_V\propto \Omega^{1/2}$.

The similar procedure gives the same DoS anomaly for vortices in the $d_{xz} + i d_{yz}$ pairing state, if the vortex is oriented along the $z$-axis. The only difference is that the $d_{xz} + i d_{yz}$ states has odd chirality $N=1$, and thus the spectrum contains in addition the branch with 
$n=0$, which forms the flat band with zero energy for $|p_z|<p_0$, where $p_0$ marks the position of the Weyl point.\cite{LeeSchnyder2015} This flat band arises from the Weyl points in bulk, as it happens in vortices in chiral Weyl superfluid $^3$He-A in Fig. \ref{CoreStates} ({\it right)}).\cite{KopninSalomaa1991,Volovik1999} The concentration of all other energy levels at $p_z\rightarrow 0$ is seen in Fig. 1b of Ref.\cite{LeeSchnyder2015}. 

{\it Conclusion}:

The phenomenon of the level condensation emerges in a superfluid/superconducting system with the symmetry protected planes in momentum space at which the superconducting gap $\Delta({\bf p})=0$ (see Ref. \cite{VolovikGorkov1985} for symmetry classification of superconducting states), and if the vortex line is oriented along the normal to such a plane.
This phenomenon does not depend on whether the bulk system has odd or even chirality, i.e. on whether the energy levels obey Eq.(\ref{Caroli}) or Eq.(\ref{chiral}), since all the levels approach zero at $p_z\rightarrow 0$. This represents another consequence of the bulk-vortex correspondence:  now it provides the connection  between the line of zeroes in the bulk spectrum and the energy level condensation at the vortex. The level condensation
gives rise to anomaly in the fermionic density of states in the vortex lattice. 

The concentration of levels at $p_z\rightarrow 0$ and the corresponding anomaly in DoS come from the region outside the vortex cores. That is why the anomaly in DoS can be described using the Doppler shifted semiclassical spectrum
$E({\bf p},{\bf r})=\sqrt{\epsilon^({\bf p}+ |\Delta({\bf p})|^2} +{\bf p}\cdot{\bf v}_s({\bf r})$, where ${\bf v}_s({\bf r})$ is the superfluid velocity of the current circulation around the vortex. As is known,
in the systems with the Dirac nodal lines in the spectrum (or with the Dirac points in the 2D system)  the  Doppler shift effect gives rise to the $\sqrt{B}$ behavior of DoS.\cite{Volovik1993}
The condensation of the energy levels in the vortex represents the different description of the same effect.

This work has been done during NORDITA Scientific Program "Physics of Interfaces and Layered Structures".
I thank Vladimir Eltsov for discussions. 
The work has been supported in part by the Academy of Finland
(project no. 284594), and by the facilities of the Cryohall
infrastructure of Aalto University.

\end{document}